\documentclass[11pt]{article}
\usepackage{amsbsy}
\usepackage{amsmath}
\usepackage{amssymb}
\usepackage{graphicx}
\usepackage{dcolumn}
\usepackage{bm}
\usepackage{psfrag}
\usepackage{color}

\newcommand{\ma}[1]{\ensuremath{\mathbb{#1}}}
\newcommand{\ve}[1]{\ensuremath{\mbox{\boldmath$#1$}}}
\newcommand\nn{\nonumber}
\newcommand{\st}[1]{\small }

\begin{document}

\title{\Large\bf Analysis of the Correlation Dimension for Inertial Particles \footnote{Version accepted for publication (postprint) on Physics of Fluids {\bf 27} 073305 (2015)}}
\author{{\it  Kristian Gustavsson$^{1,2}$, Bernhard Mehlig$^2$ and
Michael Wilkinson$^3$}
\\[3 mm]
\normalsize($^1$)
Department of Physics, University of Tor Vergata, 00133 Rome, Italy\\
($^2$)
Department of Physics, G\"oteborg University, 41296 Gothenburg, Sweden\\
($^3$) Department of Mathematics and Statistics,
The Open University, \\Walton Hall, Milton Keynes, MK7 6AA, England
}
\vspace{3mm}
\date{}
\maketitle
\par
\vspace{2cm}
 \begin{center}
 {\large\bf Abstract}
 \end{center}
We obtain an implicit equation for the correlation dimension
which describes clustering of inertial particles in a complex flow
onto a fractal measure. Our general equation involves a propagator
of a nonlinear stochastic process in which the velocity
gradient of the fluid appears as additive noise.
When the long-time limit of the propagator is considered
our equation reduces to an existing large-deviation formalism, from which
it is difficult to extract concrete results. In the short-time
limit, however, our equation reduces to a solvability condition on a partial
differential equation. In the case where the inertial particles
are much denser than the fluid, we show how this approach leads to a perturbative
expansion of the correlation dimension, for which the coefficients
can be obtained exactly and in principle to any order.
We derive the perturbation series
for the correlation dimension of inertial particles suspended in
three-dimensional spatially smooth random flows with white-noise time correlations,
obtaining the first $33$ non-zero coefficients exactly.
\par
\vspace{1.6cm}
\newpage
\newpage

\section{Introduction}

In aerosols and other suspensions of microscopic bodies it may be satisfactory to
neglect hydrodynamic interactions, and to assume
that the particles move independently. It is known that small particles moving
independently in an incompressible turbulent or complex flow
may show a pronounced tendency to cluster. This occurs if the time scale for
viscous damping, $\tau_{\rm p}$, is comparable to the smallest characteristic
time scale for fluctuations in the flow, $\tau$.
Maxey \cite{Max87} proposed that these \lq inertial particles' cluster
because they are expelled from vortices by the centrifugal effect
(if they are denser than the fluid in which they are suspended; bubbles are expected to congregate in vortices).
Later, Sommerer and Ott \cite{Som93} showed that, in common with other chaotic
dynamical processes, the trajectories of particles advected on compressible surface flows
approach a fractal attractor
 (Ott \cite{Ott02} gives a good introduction to the role of fractals in dynamical systems).
Numerical experiments by Bec \cite{Bec03} confirmed that fractal clustering
is  observed for inertial particles in incompressible flows, just as in the compressible
surface flows considered in \cite{Som93}.

This clustering is of fundamental
importance to understanding the effect of turbulence on aerosols, because of
its potential relevance to the coalescence of cloud droplets into rain
\cite{Sha03}, or of dust grains into planetary precursors \cite{Wil+08}.

The clustering process and its fractal dimension have been investigated numerically
in many works: \cite{Bec06} and \cite{Bec07} report state-of-the-art contributions. The
present theoretical understanding of this effect is
reviewed in \cite{Gus14a}.

The present paper is concerned with the analysis of the
correlation dimension, $D_2$, which is the most important dimension in physical
applications, but which is still quite poorly understood. The importance of the
correlation dimension arises from its direct relation to the two-point correlation function
of particles (given as equation (\ref{eq: 2.3}) below), which enters in theories for
collision processes \cite{Sun+97,And07} and light scattering \cite{Sin89}.

Our approach gives an implicit equation for the correlation dimension,
in terms of a propagator for a nonlinear stochastic process in which components
of the velocity gradient tensor appear as additive noise. In the limit as the propagation
time approaches zero, our equation becomes a solvability condition for a
linear partial differential equation. We analyse this system
using perturbation theory. We obtain a series expansion for the correlation
dimension of the particle distribution in powers of a dimensionless parameter which
measures the importance of inertial effects.

We briefly review the state of the theoretical knowledge concerning
the clustering of particles. Maxey's original work \cite{Max87} proposed that
the particles (which we assume are much denser than the fluid) are expelled by centrifugal
forces from vortices in the fluid, but that this effect can only be effective
when the motion of the particles relative to the fluid is neither too lightly
damped nor too heavily damped. The damping is characterised by a dimensionless
number termed the Stokes number, defined by
\begin{equation}
\label{eq: 1}
{\rm St}=\frac{1}{\gamma \tau}
\end{equation}
where $\tau$ is a characteristic time scale of the fluid flow, and where $\gamma=1/\tau_{\rm p}$ is the rate
constant for damping the motion of the particles relative to the fluid. He also showed that, when
inertial effects are weak, the particle velocity may be approximated by an effective velocity field which
has a compressible component. Simulations do show that particles have
lower density in regions of high vorticity, see \cite{Wan93}.

The particle distribution has clustering properties which are much more significant
than the instantaneous negative correlation between density and vorticity. The particles
approach a fractal measure. This can be characterised in a variety of ways, but the approach
which is most easily understood and most fundamental to physical applications is to consider
the number of particles ${\cal N}$ inside a ball of radius $\delta r$ centred on a randomly
selected test particle. For sufficiently small values of $\delta r$, the average of this quantity has a
power-law dependence upon $\delta r$ with exponent denoted by $D_2$:
\begin{equation}
\label{eq: 1.2}
\langle {\cal N}(\delta r)\rangle\sim \delta r^{D_2}\,.
\end{equation}
Throughout this paper the expectation value of $X$ is denoted by $\langle X\rangle$. The exponent
$D_2$ is termed the correlation dimension of the particle distribution \cite{Ott02}.
The fractal dimension of particle clusters has been investigated numerically, and it has been
confirmed that the fractal dimension in turbulent flows is significantly less than the space dimension only
when the Stokes number is of order unity (see for example Fig.~2 in \cite{Hog01},
and Fig.~1 in \cite{Bec07}).

However, a theoretical analysis leading to quantitative results concerning the dependence of
the dimension $D_2$ upon the Stokes number is lacking. There
are a few works in which analytical results on the correlation dimension have
been obtained. Most of the literature has discussed the relation between the Renyi dimensions
and the statistics of the finite-time Lyapunov exponent: these relationships were established
by Grassberger and Procaccia \cite{Gra84} (see also \cite{Tel87}), and are reviewed in the book
by Ott \cite{Ott02}. Usually the finite-time Lyapunov exponent can only be investigated
numerically, but its statistics can be obtained for the Kraichnan model \cite{Kra68} in which a particle
is advected in a velocity field with white-noise temporal correlations: Falkovich {\sl et al} \cite{Fal+01}
discussed the calculation of the Renyi dimensions for the Kraichnan model.
The first analytical studies on the correlation dimension for inertial particles was made
by Bec {\sl et al} \cite{Bec08}, who considered a velocity field which has white-noise temporal
correlations. Their method yields the first two terms of the series expansion of $D_2$, but
it seems to be very difficult to extend to higher orders (and the second-order
coefficient in \cite{Bec08} appears to be incorrect).
In \cite{Wil+10} we described a new method which related
the correlation dimension to the solution of a partial differential equation.
It was shown that the series expansion of the solution to this equation can be automated,
so that coefficients of arbitrary order are obtained by repeated
application of a system of annihilation and creation operators.
In this way the coefficients in a series expansion of the correlation dimension
of inertial particles in two-dimensional random flows was obtained.
Gustavsson and Mehlig \cite{Gus11a} used a different technique to
compute the correlation dimension for a random-flow model in one
dimension where the correlation dimension could be treated as a small parameter.
In a series of papers Zaichik and Alipchenkov \cite{Zai+03,Zai+07,Zai+09}
developed an approach to calculating the clustering and collision rates
of particles in a turbulent flow which combines empirical data on turbulence,
with a stability analysis of the dispersion of particles.

In this paper we describe a general principle (section \ref{sec: 2}) for calculating the
correlation dimension, based on the invariance of the distribution of
small separations under dilations, corresponding to translations in logarithmic variables.
In section \ref{sec: 3} we show how this principle can be expressed in terms of a
time-propagator. We show that a large-time
expansion of the propagator gives a set of equations closely related to equations
derived from a large-deviation principle  -- discussed in \cite{Gra84,Tel87,Ott02}.
We also show how an approximate expression for $D_2$ can be
recovered from the large deviation formalism, but it is difficult to extend this because
of the intractability of determining the entropy function of the large deviations
of the Lyapunov exponent. A short-time expansion of the propagator, by contrast, yields a partial
differential equation involving $D_2$ which is more amenable to analysis.
This approach was previously outlined in  \cite{Wil+10}.
Here, in section \ref{sec:model}, we apply the method to a white-noise random-flow model
in three spatial dimensions, developing a perturbation theory for $D_2$ in
section \ref{sec:pt}. Because the correlation time of the flow is $\tau=0$ for our
model flow, the Stokes number is not defined for our model. However, our perturbation
parameter, $\epsilon$,  plays a role which is analogous to ${\rm St}$. The relation between
$\epsilon$ and ${\rm St}$ is discussed carefully in \cite{Wil07}, where it is argued that
$\epsilon^2\propto {\rm St}$.
The perturbation series is divergent and the methods used to extract finite
results are discussed in Section \ref{sec:results}.
Section \ref{sec:conc} contains our conclusions and discusses possible extensions
of this work.

\section{The correlation dimension}
\label{sec: 2}

In Secs.~\ref{sec: 2} and \ref{sec: 3}, we define the correlation dimension
and discuss several distinct but
interconnected approaches to calculating it. What these approaches have in common is that they
use a dynamical variable, $Z_1(t)$, which is derived from the linearised equation of motion. The
statistics of $Z_1(t)$ are also closely related to the leading Lyapunov exponent.
Several different probability density functions must be introduced. We denote the probability
density function (PDF) of a quantity $X$ by a function $\rho_X$, so that the probability
element for $X$ to lie in the interval $[X,X+{\rm d}X]$ is ${\rm d}P=\rho_X(X){\rm d}X$.
The expectation value of $X$ is denoted by $\langle X\rangle$.

The correlation dimension $D_2$ is defined in terms of the expected number
$\langle {\cal N}(\delta r)\rangle$ of particles inside a ball of radius $\delta r$ surrounding a test particle:
\begin{equation}
\label{eq: 2.1}
D_2=\lim_{\delta r \to 0}\frac{{\rm ln}\langle{\cal N}(\delta r)\rangle}{{\rm ln}(\delta r)}
\end{equation}
so that
\begin{equation}
\label{eq: 2.2}
\langle{\cal N}(\delta r)\rangle\sim  \delta r^{D_2}
\end{equation}
which is the volume element of a ball in $D_2$ dimensions. If $D_2=d$ (where $d$ is the dimensionality of space) there is no clustering. The probability density $\rho(\delta r)$ for a particle to have another particle at small distance
$\delta r$ is
\begin{equation}
\label{eq: 2.3}
\rho(\delta r) = \frac{{\rm d}\langle {\cal N}(\delta r)\rangle}{{\rm d} \delta r} \sim \delta r^{D_2-1}
\ .
\end{equation}
Note that this quantity is the \lq two-point correlation function' which plays
an important role in physical kinetics \cite{Sun+97,And07} and scattering theory \cite{Sin89}.

\subsection{Logarithmic separation dynamics}
\label{sec: 2.1}

It is not immediately clear why the limit in equation (\ref{eq: 2.1}) should exist.
In this paper we show why it does, and how to extract information about $D_2$
by considering a quantity $Z_1(t)$ defined by
\begin{equation}
\label{eq: 2.4}
\frac{\delta \dot r}{\delta r}=Z_1\,.
\end{equation}
Here $\delta\dot{r}$ denotes the time derivative of $\delta r$, and
$Z_1$ is the logarithmic derivative of $\delta r$. We also consider the variable
\begin{equation}
\label{eq: 2.5}
Y(t)={\rm ln}\,\delta r(t)\,.
\end{equation}
The two variables $Y$ and $Z_1$ are related by
\begin{equation}
\label{eq: 2.6}
Y(t)=Y(0)+\int_0^t {\rm d}t'\ Z_1(t')
\,.
\end{equation}
We will argue
that, in the limit as $Y(t)\to -\infty$, the variable $Z_1$ obeys
an equation of motion which is independent of $Y$. This implies translational
invariance in the statistics of $Z_1$. Correspondingly the PDF $\rho_Y(Y)$
of $Y$ exhibits translational invariance: $\rho_Y(Y)$ and $\rho_Y(Y-Y_0)$ must be the
same function, up to a normalisation factor, for any choice of the displacement $Y_0$.
Hence
\begin{equation}
\label{eq: 2.7}
\rho_Y(Y)=C(Y_0)\rho_Y(Y-Y_0)
\end{equation}
for some choice of $C(Y_0)$. The solution of this equation is
\begin{equation}
\label{eq: 2.8}
\rho_Y(Y)=A\exp(\alpha Y)
\end{equation}
for some constant $\alpha$ and normalisation $A$. This expression is
valid only for $Y\to -\infty$, so that we must require $\alpha>0$ to give
a normalisable probability density.
Consider the corresponding PDF of $\delta r$, denoted by $\rho_{\delta r}(\delta r)$:
the element of probability is
${\rm d}P=\rho_{\delta r}(\delta r)\,{\rm d}\delta r=\rho_Y(Y)\,{\rm d}Y=A\delta r^{\alpha-1}
{\rm d}\delta r$, so that the distribution of $\delta r$ corresponding to (\ref{eq: 2.8}) is
\begin{equation}
\label{eq: 2.9}
\rho_{\delta r}(\delta r)=A\, \delta r^{\alpha-1}
\ .
\end{equation}
By comparison with (\ref{eq: 2.3}) it follows that the exponent of the
distribution of $Y$ and the correlation dimension are equal:
\begin{equation}
\label{eq: 2.10}
D_2=\alpha
\ .
\end{equation}
Thus we conclude that $D_2$ can be determined by studying the statistics of the
logarithmic derivative $Z_1=\delta \dot r/\delta r$.
Specifically, if $Z_1(t)$ is a random variable with
statistics that become independent of $Y$ as $Y\to -\infty$, then the distribution of
$Y$ is $\rho_Y(Y)\sim \exp(D_2 Y)$. So, to determine $D_2$ we need
to study the equation of motion for $Z_1(t)$ and how the statistics
of $Z_1$ determine the exponent $\alpha$.

Before going on to consider the equation of motion for $Z_1$, we remark
that the variable $Z_1(t)$ also gives information about the leading Lyapunov
exponent $\lambda$: provided the separations remain sufficiently small, we have
\begin{equation}
\label{eq: 2.11}
\lambda=\lim_{t\to \infty}\frac{1}{t} \bigg\langle{\rm ln}\left(\frac{\delta r(t)}{\delta r(0)}\right)\bigg\rangle\,.
\end{equation}
We can express this in terms of a limit of a finite-time Lyapunov exponent $\sigma(t)$:
\begin{equation}
\label{eq: 2.12}
\sigma(t)\equiv \frac{1}{t} \bigg\langle{\rm ln}\left(\frac{\delta r(t)}{\delta r(0)}\right)\bigg\rangle
=\frac{1}{t}\int_0^t {\rm d}t'\ Z_1(t')
\end{equation}
The leading Lyapunov exponent is therefore an expectation value of $Z_1(t)$:
\begin{equation}
\label{eq: 2.13}
\lambda=\lim_{t\to \infty} \sigma(t)=\langle Z_1(t)\rangle
\ .
\end{equation}

\subsection{Equation of motion for the logarithmic derivative}
\label{sec: 2.2}

We have shown that information about $D_2$ is contained in the dynamics of
the logarithmic derivative of the separation, $Z_1(t)$. To proceed further we need an equation
of motion for this quantity.  The equations of motion for a small spherical body moving
in a viscous fluid are discussed in \cite{Gat83,Max+83}. We consider the case where the density of
the body is much higher than that of the surrounding fluid. In this limit the equations of motion for the
particle position $\mbox{\boldmath$r$}(t)$ and velocity $\mbox{\boldmath$v$}(t)$ are:
\begin{equation}
\label{eq: 2.13a}
\dot{\mbox{\boldmath$r$}}=\mbox{\boldmath$v$}\,,\quad
\dot{\mbox{\boldmath$v$}}=
\gamma[\mbox{\boldmath$u$}(\mbox{\boldmath$r$}(t),t)-\mbox{\boldmath$v$}]
\ .
\end{equation}
An equation of motion for $Z_1$ is derived from the linearised equations of motion describing
a pair of particles with a separations $\delta \mbox{\boldmath$r$}$ and
$\delta \mbox{\boldmath$v$}$ in position and velocity
\begin{equation}
\label{eq: 2.14}
\delta \dot {\mbox{\boldmath$r$}}=\delta \mbox{\boldmath$v$}\,,\quad
\delta \dot {\mbox{\boldmath$v$}}=-\gamma \delta \mbox{\boldmath$v$}+\gamma {\ma E}\,\delta \mbox{\boldmath$r$}
\ .
\end{equation}
Here $\ma E$ is the matrix of flow-velocity gradients with elements $E_{ij}=\partial u_i/\partial r_j$.
From these equations we must obtain an equation of motion for $Z_1= \delta\dot r/\delta r$,
where $\delta r=|\delta \mbox{\boldmath$r$}|$.
To illustrate the approach in its simplest context, we show how
this is done for a one-dimensional model, where $x$ is the
coordinate of the particle.
In one dimension we have $\delta r=|\delta x|$, and
simple manipulation of equations (\ref{eq: 2.14}) gives
\begin{equation}
\label{eq: 2.15}
\dot Z_1=-\gamma Z_1-Z_1^2+\gamma E(t)
\end{equation}
where
\begin{equation}
\label{eq: 2.16}
E(t)=\frac{\partial u}{\partial x}(x(t),t)\,.
\end{equation}
In two or three dimensions, the variable $Z_1(t)$ is coupled to one or more additional
variables, but there are always a finite number of variables, $Z_1,Z_2,\ldots,$
which are coupled in a closed system of equations analogous to (\ref{eq: 2.15}).

The one-dimensional version of equation (\ref{eq: 2.14}) allows
particles to exchange positions, that is $\delta x$ passes through zero
while $\delta v$ remains finite. This corresponds to a
\lq caustic' singularity \cite{Wil+05} where
$Y(t)$ goes to $-\infty$ and returns, while $Z_1(t)$ goes to $-\infty$ and returns
from $+\infty$. This divergence of $Z_1$ is a special feature of the one-dimensional
version of the model and it is absent in higher dimensions. We should nevertheless
consider its effect.

The finite-time singularities give rise to a \lq tail' of the distribution of
$Y$. Consider the form of the distribution of $Y$ resulting from a
fold event in a one-dimensional system, where one phase point passes
another with a finite difference in their velocity. Because the relative velocity
has no singularity as one particle passes the other, the PDF
of the spatial separation also has no singularity. It may therefore be
approximated by a uniform distribution in the vicinity of $\delta x=0$.
The corresponding distribution for $Y$ is obtained by writing the probability element as follows
${\rm d}P=\rho_{\delta x}(\delta x){\rm d}\delta x=\rho_Y(Y){\rm d}Y$.
Hence
\begin{equation}
\label{eq: 2.17}
\rho_Y(Y)\sim {\rm const.}\times\frac{{\rm d}\delta x}{{\rm d}Y}\sim \exp(Y)
\end{equation}
This contribution
is negligible compared to that from the analysis of the differential
equation whenever the latter predicts $\alpha<1$.
The contribution from the folding events is therefore smaller
than that due to fractal clustering whenever $D_2<1$. This condition
is never violated in one dimension \cite{Gus11a}. In higher dimensions the
equation analogous to (\ref{eq: 2.13}) does not have finite-time singularities,
although there are caustic singularities where volume elements vanish
\cite{Wil+05,Wil07}.

\section{Markovian approximations}
\label{sec: 3}

We wish to use information about statistics of $Z_1(t)$ to determine $D_2=\alpha$.
The most practicable approach is to use a Markovian assumption, where
the future development of a system can be assumed to be independent of
its past history. In the present context, we assume
that future evolution of $Y(t)$ is determined by its current
value, and by the current value of $Z_1$. We therefore consider a joint
PDF of $Y$ and of $Z_1$. Given $Y$ and $Z_1$, let $K(\Delta Y,Z_1,Z_1',t)$ be the
PDF for $Y$ to increment by $\Delta Y$ and for $Z_1$ to reach $Z_1'$ after time $t$.
The joint PDF of $Y$ and $Z_1$ evolves according to
\begin{equation}
\label{eq: 3.1}
\rho_{Y,Z_1}(Y,Z_1,t)=\int_{-\infty}^\infty {\rm d}\Delta Y\int_{-\infty}^\infty {\rm d}Z_1'\
K(\Delta Y,Z_1',Z_1,\Delta t)\, \rho_{Y,Z_1}(Y-\Delta Y,Z_1',t-\Delta t)
\ .
\end{equation}
The steady-state probability density is expected to be a product
\begin{equation}
\label{eq: 3.2}
\rho_{Y,Z_1}(Y,Z_1)=\rho_{Z_1}(Z_1)\exp(\alpha Y)
\end{equation}
where the distribution $\rho_{Z_1}(Z_1)$ will be discussed shortly.
Because equation (\ref{eq: 3.1}) is derived by linearisation of the
equations of motion, equation (\ref{eq: 3.2}) is valid in the limit as $Y\to -\infty$.
In order for the distribution to be normalisable, we require that $\rho_{Y,Z_1}$ approaches
zero sufficiently rapidly as $Y\to -\infty$, implying that $\alpha>0$.

Inserting (\ref{eq: 3.2}) into (\ref{eq: 3.1}), the steady-state distribution $\rho_{Z_1}(Z_1)$ and
the exponent $\alpha$ must satisfy an integral equation
\begin{equation}
\label{eq: 3.3}
\rho_{Z_1}(Z_1)=\int_{-\infty}^\infty {\rm d}\Delta Y\int_{-\infty}^\infty {\rm d}Z_1'\
K(\Delta Y,Z_1',Z_1,\Delta t)\, \rho_{Z_1}(Z_1')\exp(-\alpha\Delta Y)
\end{equation}
which is valid for all $\Delta t$.

Consider the distribution $\rho_{Z_1}(Z_1)$ in (\ref{eq: 3.2}).
It might be expected that this is the same as the distribution of $Z_1(t)$ obtained from
equation (\ref{eq: 2.15}) or its multi-dimensional generalisation. We term this
distribution $\rho_0(Z_1)$. However, the distribution
$\rho_{Z_1}(Z_1)$ differs from $\rho_0(Z_1)$
because it is conditioned upon being at a particular value of $Y$ \cite{Wil+10}. If $\alpha\ne 0$,
particles reaching a negative value of $Z_1$ have recently arrived from a larger value of $Y$,
where the probability density is larger. This implies that the distributions are different, and
moreover that the distribution $\rho_{Z_1}(Z_1)$ has a smaller mean value than $\rho_0(Z_1)$.

Now consider the application of this equation in two limiting cases.

\subsection{Short propagation time}
\label{sec: 3.1}

Consider the limit $\Delta t\to 0$ in (\ref{eq: 3.3}). In this limit the
structure of the propagator can be simplified, because $\Delta Y=Z\Delta t+O(\Delta t^2)$.
This implies that one of the integrals can be eliminated from (\ref{eq: 3.3}), and we
may write
\begin{equation}
\label{eq: 3.4}
\rho_{Z_1}(Z_1)=\int_{-\infty}^\infty {\rm d}Z_1'\
{\cal U}(Z'_1,Z_1,\Delta t)\, \rho_{Z_1}(Z_1')\exp(-\alpha Z_1'\Delta t)
\end{equation}
where ${\cal U}(Z_1',Z_1,\Delta t)$ is the propagator for the random
process $Z_1(t)$ with equation of motion (\ref{eq: 2.15}) (or its
higher-dimensional generalisation) to reach $Z_1$
from $Z_1'$ in time $\Delta t$.

If a Markovian approximation is valid in the limit $\Delta t\to 0$, we have
continuous-time Markov process for (\ref{eq: 3.3}), and the probability
density $\rho_{Z_1}(Z_1,t)$ obeys a Fokker-Planck equation \cite{vKa81},
where the evolution kernel ${\cal U}(Z_1',Z_1,t)$ is generated by a Fokker-Planck
operator $\hat {\cal F}$:
\begin{equation}
\label{eq: 3.5}
\frac{\partial \rho_{Z_1}}{\partial t}=\hat {\cal F}\rho_{Z_1} \ .
\end{equation}
We can represent functions as vectors using Dirac notation, so that
(\ref{eq: 3.5}) is notated as follows:
\begin{equation}
\label{eq: 3.6}
\partial_t \vert \rho_{Z_1})=\hat {\cal F}\, \vert \rho_{Z_1})
\ .
\end{equation}
For small values of $\Delta t$ the action of the propagator kernel
can then be approximated by $\hat {\cal U}(\Delta t)=\hat {\cal I} +\hat {\cal F}\Delta t+O(\Delta t^2)$,
where $\hat {\cal I}$ is an identity operator, that is for a function $f(Z_1)$ represented
by a vector $|f)$, we have
\begin{equation}
\label{eq: 3.7}
\hat {\cal U}(\Delta t)|f)\equiv \int_{-\infty}^\infty {\rm d}Z_1'\
{\cal U}(Z_1',Z_1,\Delta t)f(Z_1')=|f)+\hat {\cal F} |f)\Delta t+O(\Delta t^2)
\ .
\end{equation}
For small values of $\Delta t$ equation (\ref{eq: 3.4}) then reduces
to
\begin{equation}
\label{eq: 3.7a}
\rho_{Z_1}(Z_1)=\exp(-\alpha Z_1\Delta t)\rho_{Z_1}(Z_1)+\Delta t \hat {\cal F}\rho_{Z_1}(Z_1)
+O(\Delta t^2)\,.
\end{equation}
In the limit as $\Delta t\to 0$, this relation implies the condition
\begin{equation}
\label{eq: 3.8}
[\hat {\cal F}-\alpha Z_1]\, \rho_{Z_1}(Z_1)=0
\end{equation}
which is a partial differential equation for $\rho_{Z_1}(Z_1)$ and $\alpha$.

At this stage it is useful to consider a concrete example. The one-dimensional
model equation of motion for $Z_1$, Eq. (\ref{eq: 2.15}), can be regarded as a stochastic
differential equation, in which the velocity gradient $E(t)$ is a random element. If the correlation
time of $E(t)$ is sufficiently small, a Markovian approximation is justified, and $E(t)$ can be
replaced by a multiple of a white noise signal, $\eta(t)$, which has the following statistical
properties:
\begin{equation}
\label{eq: 3.8a}
\langle \eta(t)\rangle=0
\ ,\ \ \ \
\langle \eta(t)\eta(t')\rangle=\delta(t-t')
\ .
\end{equation}
The equation of motion for $Z_1$ is replaced by
\begin{equation}
\label{eq: 3.8b}
\dot Z_1=-\gamma Z_1-Z_1^2+\sqrt{2{\cal D}}\eta(t)
\end{equation}
where the diffusion coefficient is
\begin{equation}
\label{eq: 3.8c}
{\cal D}=\frac{\gamma^2}{2}\int_{-\infty}^\infty {\rm d}t\ \langle E(t)E(0)\rangle
\ .
\end{equation}
The Fokker-Planck operator corresponding to the Langevin equation
(\ref{eq: 3.8b}) is \cite{vKa81}:
\begin{equation}
\label{eq: 3.8d}
\hat {\cal F}=(\gamma Z_1+Z_1^2)\frac{\partial}{\partial Z_1}+{\cal D}\frac{\partial^2}{\partial Z_1^2}
\end{equation}
so that for the one-dimensional model equation (\ref{eq: 3.8}) reduces to an ordinary
differential equation
\begin{equation}
\label{eq: 3.8e}
\frac{{\rm d}}{{\rm d}Z_1}\left[(\gamma Z_1+Z_1^2)\rho_{Z_1}(Z_1)+\frac{{\rm d}\rho_{Z_1}}{{\rm d}Z_1}(Z_1)\right]-
\alpha\, Z_1\rho_{Z_1}(Z_1)=0
\ .
\end{equation}
We require normalisable solutions $\rho_{Z_1}(Z_1)$, which only exist for particular
values of $\alpha$. (Later, we give a prescription leading to a unique series
solution of this equation).
Upon integrating over space, and using the fact that $\hat {\cal F}$ is a divergence, we have
\begin{equation}
\label{eq: 3.9}
\int_{-\infty}^\infty {\rm d}Z_1\ Z_1\,\rho_{Z_1}(Z_1)=\langle Z_1\rangle=0
\end{equation}
The equations (\ref{eq: 3.8}) and (\ref{eq: 3.9}) together constitute
a new and exact method for determining $D_2=\alpha$, in two steps.  First equation (\ref{eq: 3.8}) is
solved to determine a one-parameter family of solutions. Second, the correct
value of $D_2$ is determined by finding the value of $\alpha$ for which the mean value
of $Z_1$ is zero \cite{Wil+10}.

This approach has the attractive feature that it involves the analysis of differential equations,
which are susceptible to many types of mathematical techniques.

\subsection{Long-time propagation}
\label{sec: 3.2}

In the long-time limit we expect that a Markovian approximation is always valid.
In this limit the propagator is expected to \lq forget' the initial distribution, so that
\begin{equation}
\label{eq: 3.10}
K(\Delta Y,Z_1',Z_1,\Delta t)=\rho_{Z_1}(Z_1')\, \rho_{\Delta Y}(\Delta Y,\Delta t)
\end{equation}
independent of $Z_1$, where $\rho_{\Delta Y}(\Delta Y,\Delta t)$ is the probability
of a displacement  $\Delta Y$ in time $\Delta t$.

We now apply the large-deviation principle \cite{Fre+65,Tou09} to the statistics of $\Delta Y$.
This principle concerns the statistics of time averages such
as the finite-time Lyapunov exponent $\sigma(t)=\Delta Y/t$, equation (\ref{eq: 2.12}).
It is expected that the tails of the distribution $\rho_{\Delta Y}(\Delta Y)$ satisfy
\begin{equation}
\label{eq: 3.11}
\rho_{\Delta Y} (\Delta Y,t)\sim \exp[-tI(\Delta Y/t)]
\end{equation}
for some function $I(\sigma)$, which is termed the \lq entropy function' in the literature
on large-deviation theory.

The displacement is $\Delta Y=\sigma t$. Changing the variable of integration in
(\ref{eq: 3.3}) from $\Delta Y$ to $\sigma$, we obtain
\begin{equation}
\label{eq: 3.12}
\rho_{Z_1}(Z_1)=t\int_{-\infty}^\infty {\rm d}\sigma \int_{-\infty}^\infty {\rm d}Z_1'\
\rho_{Z_1}(Z_1')\rho_{Z_1}(Z_1)\exp[-t(I(\sigma)+\alpha\sigma)]
\ .
\end{equation}
Assuming that $\rho_{Z_1}(Z_1)$ is a normalised distribution, this gives
\begin{equation}
\label{eq: 3.13}
1=t\int_{-\infty}^\infty {\rm d}\sigma\
\exp[-t(I(\sigma)+\alpha \sigma)]
\ .
\end{equation}
This integral is an implicit relation between $\alpha $ and the large
deviation function $I(\sigma)$. The integral is estimated using the Laplace
principle: in the limit as $t\to \infty$, the integral is estimated by
determining the value of the integrand at its maximum.
The maximum is at position $\sigma^\ast$ determined by the condition
\begin{equation}
\label{eq: 3.15}
I'(\sigma^\ast)=-\alpha
\end{equation}
and the integral is estimated as $t\exp[-t(I(\sigma^\ast)+\alpha\sigma^\ast)]
\sim 1$, so that
\begin{equation}
\label{eq: 3.16}
I(\sigma^\ast)+\alpha \sigma^\ast=0
\ .
\end{equation}
These equations, (\ref{eq: 3.15}) and (\ref{eq: 3.16}) can, in principle, be
solved to determine $\alpha=D_2$. Similar approaches
are discussed in \cite{Gra84,Tel87,Ott02}. The difficulty lies in determining the entropy
function, $I(\sigma)$.

\subsection{An approximate expression for $D_2$}
\label{sec: 3.3}

Before exploring the applications of equation (\ref{eq: 3.8}) in
greater depth, we describe an approximate expression for $D_2$, previously
discussed in \cite{Wil+12}, which is asymptotically correct in the limit as $D_2\to 0$. Because of its simplicity, it is a natural benchmark
against which other approaches can be compared.

We observe that the variable $Y$ has diffusive fluctuations
\begin{equation}
\label{eq: 3.3.1}
\langle (\Delta Y(t)-\lambda t)^2\rangle=2{\cal D}_Y\Delta t
\end{equation}
with diffusion coefficient
\begin{equation}
\label{eq: 3.3.2}
{\cal D}_Y=\tfrac{1}{2}\int_{-\infty}^\infty {\rm d}t\ \big[\langle Z_1(t)Z_1(0)\rangle-\langle Z_1\rangle^2\big]
\ .
\end{equation}
On time scales which are large compared to the correlation time of $Z_1(t)$ we expect that
the probability density $P(Y,t)$ satisfies a  Fokker-Planck equation
\begin{equation}
\label{eq: 3.3.3}
\frac{\partial P}{\partial t}=-\frac{\partial}{\partial Y}(vP)+\frac{\partial^2}{\partial Y^2}({\cal D}_YP)
\ .
\end{equation}
The drift velocity $v$ and diffusion coefficient ${\cal D}_Y$ are defined by the relations
\begin{equation}
\label{eq: 3.3.4}
v=\frac{\langle\delta Y\rangle}{\delta t}\ ,\ \ \ {\cal D}_Y=\frac{\langle\delta Y^2\rangle}{2\delta t}
\ .
\end{equation}
When $v$ and ${\cal D}_Y$ are constant, this equation has an exponential solution
\begin{equation}
\label{eq: 3.3.5}
P=A\exp\left(\frac{v}{{\cal D}_Y} Y\right)
\ .
\end{equation}
Noting that the drift velocity $v$ is equal to the Lyapunov exponent,
$\langle Z_1\rangle=\lambda$, comparison with (\ref{eq: 2.8}) and (\ref{eq: 2.10}) implies that
\begin{equation}
\label{eq: 3.3.6}
D_2=\alpha=\frac{\lambda}{{\cal D}_Y}
\ .
\end{equation}
This approximation is only valid when $\lambda>0$, because no
normalisable solution can be constructed if $P(Y)$ is diverging as $Y\to -\infty$.

The use of the Fokker-Planck equation is only justified when the gradient of $P(Y,t)$ is sufficiently
small. The condition is that $\partial P/\partial Y$ should be small compared to $1/\delta Y_0$, where
$\delta Y_0$ is the scale over which $Y$ varies during its correlation time. The condition for the validity
of (\ref{eq: 3.3.6}) is therefore $\langle Z_1\rangle/\langle \vert Z_1\vert \rangle \ll 1$, which is equivalent to
$D_2\ll 1$.

Consider how equation (\ref{eq: 3.3.6}) relates to the long-time limit of the
propagator. The statistics of the displacement $Y(t)$ are directly related to
the finite-time Lyapunov exponent: $\Delta Y(t)=\sigma(t)$. The variance of $\sigma(t)$
is $2{\cal D}_Yt$. In the case where $I(\sigma)$ can be
adequately approximated by a quadratic function, we see that $I(\sigma)$ may be approximated
by
\begin{equation}
\label{eq: 3.3.7}
I(\sigma)=\frac{(\sigma-\lambda)^2}{4{\cal D}_Y}
\ .
\end{equation}
Using this approximation in (\ref{eq: 3.15}) and (\ref{eq: 3.16}) we recover Eq.~(\ref{eq: 3.3.6}).

\section{Three-dimensional model}
\label{sec:model}

In this section we consider how to compute the
correlation dimension for inertial particles
suspended in a three-dimensional flow.
In order to make it possible to perform the
analysis, we consider particles in a random
velocity field with known statistical properties.
This approach has been successful in modelling
the Lyapunov exponents of particles in turbulent
flows: the leading Lyapunov exponent
was obtained in \cite{Meh05}, and all three Lyapunov exponents for the
spatial separation of particles in \cite{Wil07}, showing
excellent agreement with the numerical simulations
of particles in turbulent flows described by Bec \cite{Bec06}.  Here we
build upon the results of these earlier calculations by analysing
the correlation dimension for the same random-flow model.

The flow underlying turbulent aerosols is usually incompressible,
$\ve \nabla \cdot \ve u = 0$. But in order
to analyse the properties of the perturbation theory employed
in this paper, it is of interest to also consider partially compressible flows.
We use the following decomposition of the flow velocity
into solenoidal and potential  contributions
\begin{equation}
\label{eq:ansatz}
\ve u=C_3\,(\nabla\wedge\ve A+\beta\nabla\psi)
\end{equation}
where $C_3$ is a constant.
This model was used in \cite{Meh05}
to compute the maximal Lyapunov exponent of inertial
particles in random, partially compressible flows.
The parameter $\beta$ determines the relative magnitude
of the potential  and solenoidal contributions.
A convenient measure of the relative importance
of these two contributions is
\begin{equation}
\label{eq:compr}
\Gamma = \frac{4+\beta^2}{2 + 3\beta^2}\,.
\end{equation}
Since the parameter $\beta$ assumes values between zero and infinity,
we have that ${1\over 3} \leq \Gamma \leq 2$. The case $\Gamma=2$
corresponds to solenoidal flow ($\beta =0$). For $\Gamma = {1\over  3}$,
by contrast, the flow is purely potential ($\beta \rightarrow \infty$).
A special case of interest discussed below corresponds to $\Gamma=1$,
where the solenoidal and potential contributions are of equal strengths.

We take the components of $\ve A$ and $\psi = A_0$
to be Gaussian homogeneous isotropic random functions with zero mean values
and correlation functions
\begin{equation}
\langle A_i(\ve r,t)A_j(\ve r',t')\rangle=
\delta_{ij} \,C(|\ve r-\ve r'|,|t-t'|)
\ .
\end{equation}
The correlation function $C$ is assumed to decay to zero
for spatial separations much larger than the correlation length $\eta$ of the
flow, and for time differences much larger than the correlation time $\tau$.
The typical fluctuation size of the flow is denoted by $\langle \ve u^2 \rangle= u_0^2$.
This implies that the normalisation constant in (\ref{eq:ansatz}) must be chosen as:
\begin{equation}
C_3^2=u_0^2\big [3(2+\beta^2)\,|C''(0,0)|\,\big]^{-1}\,.
\end{equation}
Following the approach in \cite{Meh05} and \cite{Wil07}
we analyse this model in the \lq white noise' limit $\tau\to 0$,
which justifies the use of the Markovian approximation considered
in section \ref{sec: 3.1}. The fluctuations of the velocity gradients ${\ma E}(t)$ are
characterised by specifying a set of diffusion coefficients, analogous
to equation (\ref{eq: 3.8c}). The diffusion coefficients are expressed in terms of the correlation functions
of the elements of $\ma E$:
 \begin{equation}
 {\cal D}_{ii}=\frac{\gamma^2}{2}
 \int_{-\infty}^\infty {\rm d}t\ \langle E_{i1}(t)E_{i1}(0)\rangle
 \end{equation}
(the factor of $\gamma^2$ is a consequence of the fact that $E(t)$
is multiplied by $\gamma$ in (\ref{eq: 2.15})).
There are some technical complications involved in calculating the
Fokker-Planck operator appearing in (\ref{eq: 3.8}), which were
discussed in detail in \cite{Meh05}. We can read off the Fokker-Planck
operator from the results in that paper. The version of (\ref{eq: 3.8})
which is applicable to our model is
 \begin{align}
 \frac{\partial \rho}{\partial t}&= \frac{\partial}{\partial Z_1}[(\gamma Z_1+Z_1^2-Z_2^2-Z_3^2)\rho ]
 +{\cal D}_{11}\frac{\partial^2 \rho}{\partial Z_1^2}
 \nn \\
& +\frac{\partial}{\partial Z_2}[(\gamma Z_2+2Z_1Z_2) \rho]
 +{\cal D}_{22}\frac{\partial^2 \rho}{\partial Z_2^2}\nn\\
 &+\frac{\partial}{\partial Z_3}[(\gamma Z_3+2Z_1Z_3) \rho]
 +{\cal D}_{33}\frac{\partial^2 \rho}{\partial Z_3^2}
 -\alpha Z_1 \rho
 \,.
 \label{eq:FP_Z}
 \end{align}
The steady-state form of this equation is analogous to the one-dimensional equation (\ref{eq: 3.8e}).
In three dimensions the condition (\ref{eq: 3.9}) takes the form
\begin{equation}
\int_{-\infty}^\infty {\rm d}Z_1 \int_{-\infty}^\infty {\rm d}Z_2 \int_{-\infty}^\infty {\rm d}Z_3\ Z_1\ \rho (Z_1,Z_2,Z_3)\equiv \langle Z_1\rangle=0
\ .
\label{eq:BC}
\end{equation}
The correlation dimension (equal to $\alpha$) is obtained by finding a value of $\alpha$ for which a normalisable solution of (\ref{eq:FP_Z}) can be obtained for which the mean value of $Z_1$ is zero.
The equations (\ref{eq:FP_Z}) and (\ref{eq:BC}) together constitute an exact method for determining the correlation dimension in the white-noise limit.

\section{Perturbation theory}
\label{sec:pt}

Here we derive a perturbation expansion for the correlation dimension.
It is convenient to introduce dimensionless variables:
\begin{equation}
x_i=\sqrt{\gamma/{\cal D}_{ii}}\,Z_i\,.
\end{equation}
The expansion parameter of the perturbation expansion is given by
$\epsilon$ where
\begin{equation}
\label{eq:defeps}
\epsilon^2 =\frac{{\cal D}_{11}}{\gamma^3}\,.
\end{equation}
Because $\epsilon \to 0$ in the over-damped limit, this perturbation parameter
plays a role which is analogous to the Stokes number. The connection between
$\epsilon$ and ${\rm St}$ is discussed in detail in \cite{Wil07}.
We denote the
joint probability density of $x_1,\ldots,x_3$
in the steady state by $P(x_1,x_2,x_3)$. It follows from Eq.~(\ref{eq:FP_Z})
that $P$ satisfies the equation:
\begin{eqnarray}
0&=&\hat{\cal  F}\,P\equiv\frac{\partial}{\partial x_1}[(x_1+\epsilon(x_1^2-\Gamma (x_2^2+x_3^2)))P]\nn\\
&+&\frac{\partial}{\partial x_2}[x_2(1+2\epsilon x_1)P]
+\frac{\partial}{\partial x_3}[x_3(1+2\epsilon x_1)P]\nonumber\\
&&+\frac{\partial^2 P}{\partial x_1^2}+\frac{\partial^2 P}{\partial x_2^2}+\frac{\partial^2 P}{\partial x_3^2}
-\epsilon \alpha x_1 P\,.
\label{eq: 15}
\end{eqnarray}
This equation defines the Fokker-Planck operator $\hat {\cal F}(\epsilon,\alpha,\Gamma)$.
Following \cite{Wil+10}, we now develop its solution as a series expansion in $\epsilon$, using a system of annihilation and
creation operators which are analogous to those used in quantum mechanics. We employ
a notation similar to the Dirac notation: a function $f(x_1,x_2,x_3)$ is denoted by a vector $|f)$.
The scalar product between two states $|f)$ and $|g)$ is given by
\begin{equation}
(f|g)\! =\!\!\!
\int_{-\infty}^\infty\!\!\!\!\!\!\!{\rm d}x_1\!\int_{-\infty}^\infty\!\!\!\!\!\!\!{\rm d}x_2\!\int_{-\infty}^\infty\!\!\!\!\!\!\!{\rm d}x_3\
f(x_1,x_2,x_3)\ g(x_1,x_2,x_3)\,.
\end{equation}
We expand both the solution $|P)$ of (\ref{eq: 15}) and the value of $\alpha$ for which the solution of this equation exists and
satisfies $\langle x_1 \rangle=0$ as power series in $\epsilon$:
\begin{equation}
\label{eq: 16}
|P ) =\sum_{k=0}^\infty \epsilon^k\, |P_k)
\ ,\ \ \
\alpha = \sum_{k=0}^\infty \epsilon^k\, \alpha_k
\ .
\end{equation}
The Fokker-Planck operator in Eq.~(\ref{eq: 15})
is written as
\begin{equation}
\label{eq: 17}
\hat {\cal F}=\hat {\cal F}_0+\epsilon (\hat {\cal G} -\alpha \hat x_1)
\end{equation}
which defines the operators $\hat {\cal F}_0$ and  $\hat {\cal G}$.
The unperturbed steady-state $|P_0)$ satisfies
\begin{equation}
\hat {\cal F}_0|P_0)=0\,.
\end{equation}
It is given by
\begin{equation}
P_0(x_1,x_2,x_3)=\frac{\exp[-(x_1^2+x_2^2+x_3^2)/2]}{(2\pi)^{3/2}}\,.
\end{equation}
Other eigenfunctions of $\hat {\cal F}_0$ are generated by creation operators $\hat a_i$ and annihilation operators $\hat b_i$:
\begin{align}
& \hat a_i =-\partial_{x_i}\,\nn\\
&  \hat b_i=\partial_{x_i}+x_i \,.
\label{eq: 18}
\end{align}
These operators generate eigenfunctions satisfying
\begin{equation}
\hat {\cal F}_0|\phi_{pnm})=-(n+m+p)|\phi_{pnm})
\end{equation}
according to the rules
\begin{align}
& \hat a_1 |\phi_{p,n,m})=|\phi_{p+1,n,m})\nn\\
& \hat b_1 |\phi_{p,n,m})=p|\phi_{p-1,n,m}) \nn\\
& \hat a_2 |\phi_{p,n,m})=|\phi_{p,n+1,m}) \nn\\
& \hat b_1 |\phi_{p,n,m})=n|\phi_{p,n-1,m}) \nn\\
& \hat a_3 |\phi_{p,n,m})=|\phi_{p,n,m+1}) \nn\\
& \hat b_1 |\phi_{p,n,m})=m|\phi_{p,n,m-1})\,.
\label{eq: 19}
\end{align}
with $|\phi_{000})=|P_0)$, normalised as a probability density.
The states $|P_k)$ in (\ref{eq: 16}) are expressed as linear combinations of the eigenfunctions $|\phi_{pnm})$:
\begin{equation}
\label{eq: 20}
|P_k)=\sum_{p=0}^\infty\sum_{n=0}^\infty \sum_{m=0}^\infty p^{(k)}_{pnm}\,|\phi_{pnm})
\ .
\end{equation}
The eigenfunctions generated by repeated applications
of $\hat a$ are neither normalised  nor do they form an orthogonal set.
This is different from earlier perturbation theories for the Lyapunov
exponent of inertial particles \cite{Meh04,Wil07}.

We first consider how the condition $\langle x_1\rangle=0$ constrains
the coefficients $p^{(k)}_{pnm}$ in (\ref{eq: 20}). Using (\ref{eq: 18}) we find:
\begin{align}
\langle x_1 \rangle & = (x_1|P)   = \sum_{k} \epsilon^k\sum_{p,n,m}
p^{(k)}_{pnm} (x_1|\phi_{pnm}) = \sum_k p^{(k)}_{100}\epsilon^k\,,
\end{align}
so that the condition $\langle x_1\rangle=0$ is satisfied by requiring that
\begin{equation}
\label{eq: 5.13a}
p^{(k)}_{100}=0
\end{equation}
for all values of $k$.
Substituting (\ref{eq: 16}) into (\ref{eq: 17}) leads to a recursion for $|P_k)$.
The term of order $\epsilon^k$ is given by
\begin{equation}
\label{eq: 22}
0=\hat {\cal F}_0 |P_k)+\hat {\cal G}|P_{k-1})-\sum_{l=0}^{k-1}\alpha_l(\hat a_1+\hat b_1)|P_{k-l-1})\,,
\end{equation}
with
\begin{align}
\hat {\cal F}_0&=-\hat a_1\hat b_1-\hat a_2\hat b_2-\hat a_3\hat b_3
\end{align}
and
\begin{align}
\hat {\cal G}&=-\hat a_1[((\hat a_1+\hat b_1)^2
 -\Gamma ((\hat a_2+\hat b_2)^2+(\hat a_3+\hat b_3)^2))]\nn\\
&-2(\hat a_1+\hat b_1)[\hat a_2(\hat a_2+\hat b_2)+\hat a_3(\hat a_3+\hat b_3)]\,.
\end{align}
Equation (\ref{eq: 22}) is a recursion for the state $|P_k)$, and the coefficient
$\alpha_{k-1}$ in terms of the states $|P_j)$ and coefficients $\alpha_j$ determined in previous iterations.
By considering the coefficient of $|\phi_{pnm})$, we obtain $p^{(k)}_{pnm}$ in terms
of the coefficients $p^{(k-1)}_{p',n',m'}$ and $\alpha_l$, with $l=0,\ldots,k-1$. In order
to determine the coefficients $\alpha_l$,
consider the case $p=n=m=0$, where the coefficient of the state $|\phi_{000})$ in
Eq.~(\ref{eq: 22}) reduces to the condition:
\begin{equation}
\label{eq:condition}
\sum_{l=0}^{k-1}\alpha_l\, p^{(k-l-1)}_{100}=0\,.
\end{equation}
This condition can be fulfilled in at least two ways.
One solution is obtained by setting all $\alpha_k$ equal to zero.
This case corresponds to calculating the Lyapunov exponent.
Using $\lambda = \langle Z_1\rangle = \sqrt{{\cal D}_{11}/\gamma}\,\,\langle x_1\rangle$ we find Eqs.~(67,68) in \cite{Meh05}.
A second possibility is to require $\langle x_1\rangle=0$
corresponding to $p^{(k)}_{100}=0$ for all values of $k$, as explained above.
This is the case relevant for calculating the correlation dimension.
Using the initial condition
\begin{equation}
p^{(0)}_{pnm}=\delta_{p0}\delta_{n0}\delta_{m0}
\end{equation}
we can iterate Eq. (\ref{eq: 22}) to determine the coefficients $p^{(k)}_{pnm}$
in terms of the $\alpha_l$. The first few non-vanishing coefficients obtained by
recursion of (\ref{eq: 22}) are listed in Table~\ref{tab:pcoeffs}.
Note that Eq.~(\ref{eq: 22}) only relates coefficients
$p^{(k)}_{pnm}$ with indices $n$ and $m$ to coefficients with indices $n'$ and $m'$ provided
that  $n'-n$ and $m'-m$ are even integers.
This implies that only coefficients with even values of $n$ and $m$ are non-zero
(see Table~\ref{tab:pcoeffs}).  Note also that if all odd-order coefficients
$\alpha_{2n+1}$ vanish, then Eq.~(\ref{eq: 22}) does not mix the parity of $p+k$
in $p^{(k)}_{pnm}$. We find that in this case,
Eq.~(\ref{eq: 22}) provides two independent recursions:
one for $p^{(k)}_{pnm}$ with even $p+k$
and the initial condition
$p^{(0)}_{2p,n,m}=\delta_{p0}\delta_{n0}\delta_{m0}$
when $k=0$, and one for $p^{(k)}_{pnm}$ with odd $p+k$ and the initial condition
$p^{(0)}_{2p+1,n,m}=0$
when $k=0$. Using the boundary condition $p^{(k)}_{100}=0$ (Eq. (\ref{eq:condition0}))
we find that indeed all odd order $\alpha_{2n+1}$ vanish, and that all coefficients with odd values
of $p+k$ vanish.  This is illustrated to the lowest order in $k$ in Table~\ref{tab:pcoeffs}:
all displayed coefficients which are of different parity in $p$ and $k$ are multiplied by $\alpha_1=-p^{(2)}_{100}=0$.

These considerations do not determine the normalisation of the distribution
$|P)$. Expanding the normalisation condition in terms of
Eqs.~(\ref{eq: 16}) and (\ref{eq: 20}) yields
\begin{equation}
\label{eq:condition0}
p^{(k)}_{000}=\delta_{k0}\,.
\end{equation}

Once the coefficients $p^{(k)}_{mnp}$ have been determined to each order $k$, we use
Eq. (\ref{eq: 5.13a}) to compute $\alpha_{k-1}$.
From Table~\ref{tab:pcoeffs} we find the first three coefficients in the expansion (\ref{eq: 16}) of $\alpha$ in powers of $\epsilon$:
\begin{equation}
\alpha_0=2\Gamma-1\,,\quad \alpha_1=0\,,\quad \alpha_2=-2(\Gamma-1)\Gamma(2\Gamma+1)\,.
\end{equation}
This gives the correlation dimension for the three-dimensional random-flow model in the white-noise limit
\begin{equation}
D_2=2\Gamma-1 -2\Gamma(\Gamma-1)(2\Gamma+1) \epsilon^2+\ldots
\end{equation}
to second order in $\epsilon$.
As mentioned in the Introduction, the two leading non-zero coefficients of this expansion for incompressible flows ($\Gamma=2$) were computed in \cite{Bec08}. The coefficient of
$\epsilon^2$ in that work differs from our result.

\section{Results and discussion}
\label{sec:results}

We use an algebraic manipulation program to obtain the series expansion of $D_2(\epsilon)$ in powers of $\epsilon$
from Eq.~(\ref{eq: 22}) to higher orders in $k$. To order $\epsilon^{10}$ the result is
\begin{align}
D_2&=2\Gamma-1
+2\Gamma(\Gamma-1)(2\Gamma+1)[
-\epsilon^2
+(-11-2\Gamma + 6\Gamma^2)\epsilon^4\nn\\
&
+\tfrac{1}{3}(-588+5\Gamma+391\Gamma^2-16\Gamma^3-144\Gamma^4)\epsilon^6
\nn\\&
+\tfrac{1}{9}(-42579+18573\Gamma+22727\Gamma^2-19284\Gamma^3-12648\Gamma^4+3464\Gamma^5+3960\Gamma^6)\epsilon^8
\nn\\&
+\tfrac{1}{27}(-3863052+3918303\Gamma+288351\Gamma^2-4153120\Gamma^3+47186\Gamma^4\nn\\ &\hspace*{5cm}+1409736\Gamma^5+277928\Gamma^6-216448\Gamma^7-117936\Gamma^8)\epsilon^{10}]
\nonumber\\&
+O(\epsilon^{12})\,.
\label{eq:D2whitenoise}
\end{align}
In special cases we have obtained expansions to higher orders.
For incompressible flows ($\Gamma=2$), for example,
we find to order $\epsilon^{28}$:
\begin{align}
D_2&=3-20\epsilon^2+180\epsilon^4-9640\epsilon^6+206940\epsilon^8\nn\\
& -16548920\epsilon^{10}+477315000\epsilon^{12}-50149424368\epsilon^{14}
\nn\\
& +1692947357004\epsilon^{16}-5614110582647928/25\epsilon^{18}\nn\\
&+ 209543657412608424/25\epsilon^{20}\nn\\
&-860424252594210743568/625\epsilon^{22} \nn\\
& +241528608428504721258888/4375\epsilon^{24}\nn\\
&-8471768050800513607578954992/765625\epsilon^{26} \nn\\
&+2514450499347358305045823304592/5359375\epsilon^{28}\nn\\
&+\ldots
\end{align}
For incompressible flows we have obtained the first $33$ non-vanishing
coefficients as fractions of integers
and the following $17$ coefficients to ten significant digits.

\begin{figure}
\includegraphics[width=12.5cm]{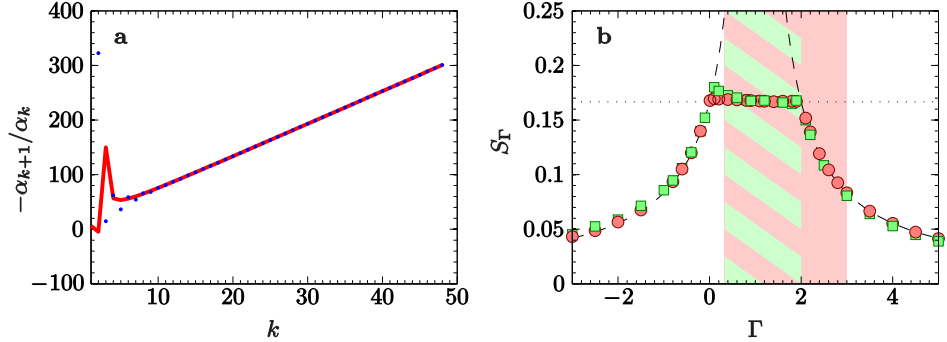}
\caption{\label{fig:1} ({\em Color online}).
{\bf a} Fit to asymptotic form (\ref{eq:action_fit}) (red solid line) for $\Gamma=3/2$ and $d=3$ in the range $25< k\le 50$.
The quotients $-\alpha_{k+1}/\alpha_k$ are shown as blue symbols.
The resulting values of the fitted parameters are $S_{\Gamma=3/2}=0.167$ $b_{\Gamma=3/2}=4.8$.
{\bf b} Shows actions obtained from the perturbation coefficients as a function of $\Gamma$, in two and in three spatial dimensions.
The first $110$ non-vanishing coefficients in two dimensions and the first
$50$ non-vanishing coefficients in three spatial dimensions are fitted to (\ref{eq:action_fit}) using a non-linear least-squares method.
The fitted actions $S_\Gamma$ are shown as symbols: red $\circ$ in two dimensions, and green $\Box$ in three dimensions.
Also shown are the curves $S=1/6$ (dotted black) and $S=1/(6|\Gamma-1|)$ (dashed black).
The coloured regions indicate the allowed ranges of the parameter $\Gamma$:
$1/3\le\Gamma\le 3$ in two spatial dimensions (red) and $1/3\le\Gamma\le 2$ in three spatial dimensions (hashed green).}
\end{figure}

The corresponding series in two spatial dimensions was derived by \cite{Wil+10}:
\begin{align}
D_2&= \Gamma-1-\Gamma(\Gamma^2-1)\epsilon^2 +\Gamma(\Gamma^2-1)(3\Gamma^2+2\Gamma-11)\epsilon^4
           +O(\epsilon^6)\,.
\label{eq:r2}
\end{align}
Iterating the recursions derived by \cite{Wil+10} we have obtained the first non-vanishing $110$ coefficients to ten significant
digits in the incompressible case ($\Gamma=3$).

The series quoted above are asymptotically divergent: they diverge but every partial sum of the series approaches $D_2$ as $\epsilon\to 0$.
Evaluating the coefficients $\alpha_k$ for a given value of $\Gamma$
shows that they grow factorially as a function of $k$:
\begin{align}
\alpha_k\sim a_\Gamma\,S_\Gamma^{-k}(k-1)!(1-b_\Gamma/k+\cdots )\,.
\label{eq:alfak_ansatz}
\end{align}
This is a typical asymptotic behaviour of the
coefficients $\alpha_k$ for large values of $k$ \cite{Dingle}.
Here the \lq action\rq~$S_\Gamma$ and the constant $b_\Gamma$ are obtained by fitting of the ansatz (\ref{eq:alfak_ansatz}) to the coefficients.
For the fit we use a non-linear least-squares method, assuming that the relative magnitude of subsequent coefficients is on the form
\begin{align}
\frac{\alpha_{k+1}}{\alpha_k}&\sim
S_\Gamma^{-1}\frac{k^2}{k+1}\frac{k+1-b_\Gamma}{k-b_\Gamma}\,.
\label{eq:action_fit}
\end{align}
Fig.~\ref{fig:1}{\bf a} illustrates the asymptotic behaviour of the
coefficients, using the case $\Gamma=3/2$ in three spatial dimensions
as an example.
The action $S_\Gamma$
extracted from fits such as the one in Fig.~\ref{fig:1}{\bf a} is shown
in Fig. \ref{fig:1}{\bf b}, in both two and three spatial dimensions.
The resulting action is found to depend upon $\Gamma$ as follows:
\begin{equation}
S_\Gamma=\min[1/6,1/(6|\Gamma-1|)]\,,
\end{equation}
in both two and three spatial dimensions.
We note that the coefficients of the perturbation
series for the maximal Lyapunov exponent
in two spatial dimensions \cite{Meh04} give rise
to the action $1/(6|\Gamma-1|)$ for all values of $\Gamma$.

We also note that the two- and three-dimensional cases shown in Fig.~\ref{fig:1}{\bf b}
differ from each other. In three dimensions the action is always
given by $1/6$ in the allowed range of $\Gamma$ (this is not the case in two spatial dimensions).
As opposed to the perturbation expansions for the Lyapunov
exponent and the two-dimensional correlation dimension, the three-dimensional perturbation expansion
for the correlation dimension is determined by one action only, $S=1/6$.

\begin{figure}
\includegraphics[width=8.0cm]{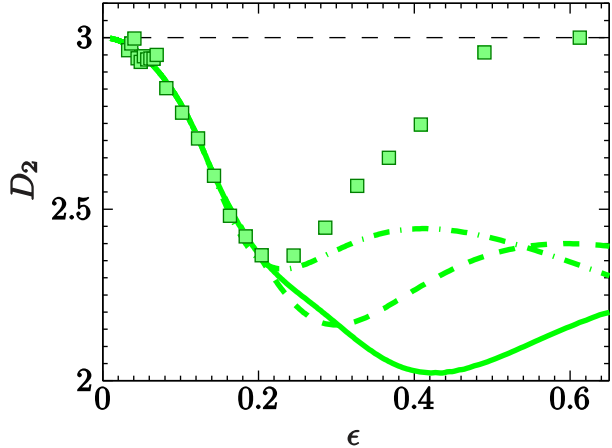}
\caption{\label{fig:2} ({\em Color online}).
Correlation dimension for the white-noise model in three spatial dimensions
for  $\Gamma = 2$
as a function of $\epsilon$.  Shown are results of direct numerical simulations of the equation
of motion (\ref{eq: 2.13a}), symbols, and results of Pad\'e{}-Borel resummations
of the perturbations series for $D_2$, of order $[4,4]$ (dash-dotted line)
 $[8,8]$ (dashed line),  $[16,16]$ (solid line).
}
\end{figure}

We have resummed the perturbation series
(\ref{eq:D2whitenoise}) and (\ref{eq:r2}) using Pad\'e-Borel resummation: to sum the series
\begin{equation}
D_2(\epsilon^2) \sim \sum_{l=0}^\infty \alpha_{2l} \epsilon^{2l}\,,
\end{equation}
consider the modified series, the so-called \lq Borel sum' (assumed to have
a finite radius of convergence due to the extra factor of $1/l!$)
\begin{equation}
B(\epsilon^2) = \sum_{l=0}^\infty \frac{\alpha_{2l}}{l!} \epsilon^{2l}
\end{equation}
Then the sum  is estimated by
\begin{equation}
\label{eq:pdbt}
D_2(\epsilon^2) = \mbox{Re} \int_C {\rm d}t \,{\rm e}^{-t} B(\epsilon^2 t)\,.
\end{equation}
The integration path $C$ is taken to be a ray in the upper right
quadrant of the complex plane. In order to perform the integral,
an approximation of the Borel sum outside its radius of convergence
is required. One possibility is to approximate $B$ by  \lq Pad\'e approximants' \cite{Ben78}
of order $[n,n]$ (or $[n,n+1]$).  For $\Gamma=2$, the Pad\'e approximations of
order $[2,2]$ and $[3,3]$ are (with $x=\epsilon^2)$:
\begin{equation}
B_{[2,2]}(x) = 3+\frac{ {\frac { 48180}{ 721}}\,{x}^{2}-20\,x } {  1+{\frac {
1671}{1442}}\,x-{\frac {649927}{8652}}\,{x}^{2} }\,,
\end{equation}
\begin{eqnarray}
B_{[3,3]}(x)= 3 +\frac{ -{\frac {728234642879}{562607103}}\,{x}^{3}+{\frac {
91933567500}{187535701}}\,{x}^{2}-20\,x }{ 1-{\frac {
7505535441}{375071402}}\,x-{\frac {99077893373}{937678505}}\,{x}^{2}+{
\frac {11726142610857}{7501428040}}\,{x}^{3}}\,.\nn
\end{eqnarray}
Higher orders are too lengthy to write down here. Fig.~\ref{fig:2}
shows the results we obtained for $D_2$ for $\Gamma=2$ in three spatial dimensions
by integrating $B_{[4,4]}$, $B_{[8,8]}$, and $B_{[16,16]}$ according to
Eq.~(\ref{eq:pdbt}). The corresponding contour $C$ in the complex $t$-plane
was chosen along a ray from the origin at angle $\pi/4$.
For small values of $\epsilon$ the results depend only negligibly on the precise choice of the contour.
Also shown are results for $D_2$ obtained by direct numerical simulations of
the equations of motion (\ref{eq: 2.7}). We observe that the Pad\'e{}-Borel resummations
converge quickly for not too large values of $\epsilon$, and we
find excellent agreement with results of direct numerical simulations of the random-flow model.

\begin{figure}
\includegraphics[width=12.5cm]{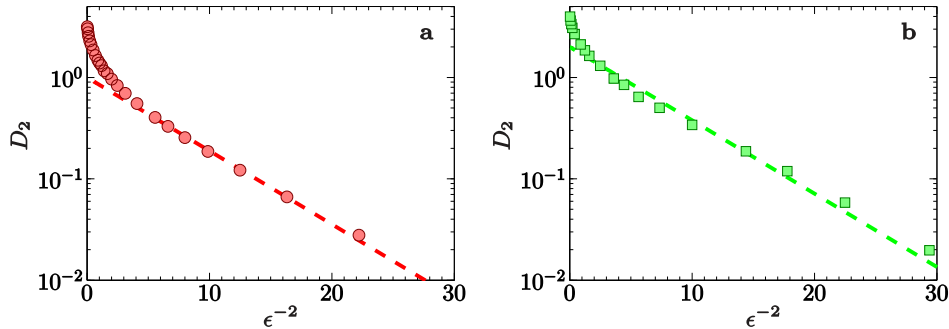}
\caption{\label{fig:4}  ({\em Color online}). Correlation dimension for $\Gamma=1$ as a function of
$\epsilon^{-2}$ in two spatial dimensions ({\bf a}) and in three spatial dimensions ({\bf b}).
Also shown is the non-analytical law (\ref{eq:causticrate})  with prefactors $A_1=1$ in
two dimensions  and $A_1=2$§ in three dimensions.
(dashed lines). }
\end{figure}

It is clear, on the other hand,  that the resummation fails for larger
values of $\epsilon$. We suspect that a non-analytical contribution
of the form $A\,\exp[-1/(6\epsilon^2)]$ is not captured and must
be added to the perturbation series. In \cite{Meh04} it is shown that a corresponding term
must to be added to the perturbation result for the maximal Lyapunov exponent.
The situation here is similar. This is most easily seen by
considering the case $\Gamma = 1$. Eq. (\ref{eq:D2whitenoise}) shows that
the first twenty  perturbation coefficients vanish for $\Gamma=1$.
We hypothesise that all coefficients vanish at $\Gamma=1$ and that
the correlation dimension
exhibits a non-analytic dependence on $\epsilon$, of the form
\begin{equation}
\label{eq:causticrate}
D_2 \sim A_1\, \exp\Big(-\frac{1}{6\epsilon^2}\Big)\,.
\end{equation}
This is shown in two and three spatial dimensions
in Fig. \ref{fig:4}.
These results complement earlier studies of the information dimension $D_1$, discussed
in detail in \cite{Wil07}: the Borel summation technique was more successful in that case,
but $D_1$ has less direct physical significance.

We conclude this section with two further comments.
First, for small values of $\epsilon$ we see that
in incompressible flows $3-D_2 \propto  \epsilon^2 \propto {\rm St}$.
This is a consequence of the fact that we considered the white-noise limit. In this limit
the fractal information dimension exhibits the same scaling \cite{Wil07}. In flows
with finite correlation time, by contrast, the correlation dimension deficit
behaves as $3-D_2 \propto {\rm St}^2$ for small Stokes numbers \cite{Bal01,Chu05,Fal04}.

Second, we note that the correlation dimension exhibits a
singularity in the advective limit ($\epsilon=0$) as the compressibility
parameter approaches $\Gamma = 1/2$, corresponding
to a path-coalescence transition where the maximal Lyapunov
exponent changes sign \cite{Meh05}. The perturbation theory
gives correct results for $\Gamma \geq 1/2$, it fails
for $\Gamma < 1/2$.

\section{Conclusions}
\label{sec:conc}

In this paper we have derived a general method for calculating the correlation dimension of
random dynamical systems, which complements DNS (direct numerical simulation) studies of
particles in turbulence \cite{Bec06,Bec07} and numerical studies
of stochastic models \cite{Zai+03,Zai+07,Zai+09}.
The method is formulated in terms of a propagator describing
the time evolution of particle separations and particle-velocity gradients.
In special cases, known methods for computing the correlation dimension
are obtained \cite{Gra84,Tel87,Ott02,Wil+12}.

A short-time expansion of the propagator  yields a solvability condition on a partial differential equation,
leading to a perturbative
expansion of the correlation dimension, for which the coefficients
can be obtained exactly and to any order. We derived the exact first $33$ coefficients
in a series expansion of the correlation dimension for inertial particles in three-dimensional
spatially smooth random flows that are white noise in time.
Related series expansions have been presented for Lyapunov exponents
of inertial particles in such flows in earlier works \cite{Meh05,Wil07,Meh04}

We have obtained accurate results for the correlation dimension of inertial particles in three-dimensional
white-noise flows by Pad\'e{}-Borel resummation of the perturbation series for not
too large values of $\epsilon$. However, for the correlation dimension $D_2$
the resummation method is not as successful as for the information dimension $D_1$,
which was considered in \cite{Wil07}. It would be desirable to develop a more direct
analytical approach to extracting information about $D_2$ from equation (\ref{eq:FP_Z}).

In a particular case, for $\Gamma=1$, we find that the perturbation coefficients vanish
and the correlation dimension exhibits a non-analytical dependence upon $\epsilon$.
We conjecture that there is a corresponding non-analytical contribution also for $\Gamma >1$.

Finally, we remark that it is possible
to extend the method presented here to treat
velocity fields with finite correlation time. This
can be achieved by considering a temporally smooth
velocity gradient obtained from a stochastic process which is driven by a white noise
signal. Numerical studies of velocity gradient statistics in turbulence
show that they have correlation functions which are well approximated
by exponentials \cite{Pum+11}. Velocity gradients of turbulent flows
can, therefore, be modelled by an Ornstein-Uhlenbeck process \cite{Orn30}, as described
in \cite{Pum+11}. The operator methods used in this present paper have been
extended to temporally smooth velocity gradients \cite{Wil09},
but some care may be required in their application and interpretation \cite{Wil09b,Gus11b}.

{\em Acknowledgements.}
Michael A. Morgan (Seattle University) helped in the initial stages of exploring
the series expansion of $D_2$ discussed in section \ref{sec:pt}.
Support from Vetenskapsr\aa{}det and the  G\"oran Gustafsson
Foundation for Research in Natural Science and Medicine is gratefully acknowledged.
K.G. acknowledges partial funding from the European Research Council under the European Community’s Seventh Framework Programme, ERC Grant Agreement N. 339032.

\newpage

\begin{table}
\mbox{}\\[1cm]
\centering
\rotatebox{90}{
\begin{tabular}{l}
\hline\hline\\[-0.25cm]
$p^{(0)}_{000}=1$
\cr
\\[-0.25cm]
\hline\\[-0.25cm]
$p^{(1)}_{100}=2\Gamma-1-\alpha_0$\,,
$p^{(1)}_{102}=p^{(1)}_{120}=\frac{\Gamma-2}{3}$\,,
$p^{(1)}_{300}=-\frac{1}{3}$
\cr
\\[-0.25cm]
\hline\\[-0.25cm]
$p^{(2)}_{002}=p^{(2)}_{020}=\frac{8\Gamma-7}{3}+\frac{\Gamma-8}{6}\alpha_0$\,,
$p^{(2)}_{022}=\frac{\Gamma-2}{3}$\,,
$p^{(2)}_{004}=p^{(2)}_{040}=\frac{\Gamma-2}{6}$\,,
$p^{(2)}_{200}=-\frac{(4\Gamma-5)(4\Gamma-3)}{6}+\frac{4\Gamma-5}{2}\alpha_0-\alpha_0^2/2$\,,
\\[0.1cm]
$p^{(2)}_{202}=p^{(2)}_{220}=\frac{(-13+17\Gamma-6\Gamma^2)}{6}+\frac{\Gamma-2}{3}\alpha_0$\,,
$p^{(2)}_{204}=p^{(2)}_{240}=-\frac{(\Gamma-2)^2}{18}$\,,
$p^{(2)}_{222}=-\frac{(\Gamma-2)^2}{9}$\,,
$p^{(2)}_{400}=\frac{4\Gamma-5}{6}-\frac{\alpha_0}{3}$\,,
$p^{(2)}_{402}=p^{(2)}_{420}=\frac{\Gamma-2}{9}$\,,
\cr
\\[-0.25cm]
\hline
\\[-0.25cm]
$p^{(3)}_{100}=-5+20 \Gamma-16 \Gamma^2+2(-5+10\Gamma-3\Gamma^2)\alpha_0+2(2\Gamma-3)\alpha_0^2-\alpha_0^3-\alpha_2$\,,
\\[0.1cm]
$p^{(3)}_{102}=p^{(3)}_{120}=\frac{-130+247 \Gamma-118 \Gamma^2}{9}+\frac{-20+19\Gamma-2\Gamma^2}{2}\alpha_0+\frac{5\Gamma-28}{18}\alpha_0^2$\,,
$p^{(3)}_{104}=p^{(3)}_{140}=\frac{-122+167 \Gamma-56 \Gamma^2}{30}+\frac{-22+13 \Gamma-\Gamma^2}{18}\alpha_0$\,,
\\[0.1cm]
$p^{(3)}_{122}=\frac{-122+167 \Gamma-56 \Gamma^2}{15}+\frac{-22+13 \Gamma-\Gamma^2}{9}\alpha_0$\,,
$p^{(3)}_{124}=p^{(3)}_{142}=-\frac{(\Gamma-2)^2}{6}$\,,
\\[0.1cm]
$p^{(3)}_{300}=\frac{-135+290 \Gamma-212 \Gamma^2+56 \Gamma^3}{18}+\frac{-67+77 \Gamma-24 \Gamma^2}{9}\alpha_0+\frac{9\Gamma-19}{9}\alpha_0^2-\frac{\alpha_0^3}{6}$\,,
\\[0.1cm]
$p^{(3)}_{302}=p^{(3)}_{320}=\frac{-715+1194 \Gamma-724 \Gamma^2+172 \Gamma^3}{90}+\frac{-65+61 \Gamma-18 \Gamma^2}{18}\alpha_0+\frac{\Gamma-2}{6}\alpha_0^2$\,,
$p^{(3)}_{304}=p^{(3)}_{340}=\frac{2}{9}(-6+9 \Gamma-5 \Gamma^2+\Gamma^3)-\frac{(\Gamma-2)^2}{18}\alpha_0$\,,
\\[0.1cm]
$p^{(3)}_{322}=\frac{4}{9}(-6+9 \Gamma-5 \Gamma^2+\Gamma^3)-1/9 (-2+\Gamma)^2\alpha_0$\,,
$p^{(3)}_{324}=p^{(3)}_{342}=\frac{(\Gamma-2)^3}{54}$\,,
\cr\\[-0.25cm]
\hline\hline
\end{tabular} }
\mbox{}\\[1cm]
\caption{\label{tab:pcoeffs} Non-zero expansion coefficients $p^{(k)}_{pnm}$ in Eq.~(\ref{eq: 20}) for $k=0,\dots,3$. Obtained
by recursive solution of Eq.~(\ref{eq: 22}).}
\end{table}

\end{document}